# Multiferroicity in $Li_{0.05}Ti_{0.02}Ni_{0.93}O$ Above Room Temperature

Jitender Kumar, Pankaj K. Pandey, and A.M. Awasthi[*]

UGC-DAE Consortium for Scientific Research, University Campus, Khandwa Road, Indore- 452 001, INDIA

## Abstract

We present direct evidence of above room temperature magneto-electricity in single-phase $Li_{0.05}Ti_{0.02}Ni_{0.93}O$ with above-ambient antiferromagnetic ordering. Temperature-hysteresis in warming/cooling heat-flow thermograms establishes a discontinuous/first-order nature of the cubic-to-rhombohedral structural change concurring the AFM transition. At $T_N = 488K$, magnetization features a sharp slope-&-curvature discontinuity and dielectric constant shows a peak-anomaly. Room temperature $P$-$E$ loop measures a large polarization traceable to the Nèel domain walls. We find *positive* room-temperature magneto-capacitance up to 9T, linear at low-fields (MC $\propto H_{low}$); the prefactor slope ($d\ln\varepsilon'/dH_{low}$) featuring a positive frequency-coefficient rises from $O(10^{-3}/T)$ at 0.1MHz to $O(10^{-1}/T)$ at 4MHz, with maximum $d\varepsilon'/dH \sim 120 T^{-1}$.

Multiferroics with two or more coexistent ferroic orders are the materials useful for the construction of four-state logic-storage devices. The coupling of electrical and magnetic orders is the key to the functional aspect of a magneto-electric (ME) multiferroic. With ever-rising demands for the modern storage devices, intensive research is focused on the multiferroic materials; unfortunately the low energy-scale (temperature, frequency, and fields'-strength) of coupling of these two orders renders very few ME's as serious device candidates.[1] Almost in all known ME's, this coupling is realized below the room temperature, except $BiFeO_3$ featuring above-ambient magnetic and electrical orderings. Although $BiFeO_3$ is a very rare ME multiferroic, having both *ferro*magnetic and *ferro*electric orders in its thin film form,[2] it too has issues with reproducivity and stability of the results, and presence of impurity phases.[3] The true room temperature single phase ME multiferroic is as elusive as the room temperature superconductor! We present here a peculiar ME character above room temperature in a simple material, obtained in single phase without extraneous effort.

---

[*] Corresponding Author e-mail: amawasthi@csr.res.in. Tel: +91 731 2463913.

A few years ago, an interesting letter was published on cupric oxide,[4] wherein AFM-induced FE is the cause of multiferroicity. Here, we report witnessing magneto-electricity in a double-doped derivative $Li_{0.05}Ti_{0.02}Ni_{0.93}O$ (LTNO) of nickel oxide, with concurring above-ambient AFM and structural transitions, and contend its placement in the extremely scarce group of *room temperature "multiferroics"*.

The parent compound NiO is a Mott-Hubbard insulator, which becomes semiconducting when doped with Li and/or Ti. $Li^{+1}$ transforms $Ni^{+2}$ into $Ni^{+3}$ for exacting the charge-neutrality, creating locally-distorted rhombohedra.[5] The motivation for synthesizing LTNO is to facilitate its dielectric characterization; inhibited in the pure polycrystalline NiO, due to its significant conductivity.[6] Thus, the perturbatively-doped LTNO specimen is supposed to have most of its other (lattice, vibrational, electronic etc.) attributes little different from the parent NiO, carrying over their features as mentioned below. In pure NiO, appearance of the super-lattice peaks (reflections) in neutron (TEM) diffraction[7] confirms the structural change at $T_N$, concurrent with the softening of the optical phonon mode.[8] A sharp transition in the thermal conductivity of NiO[9] at $T_N$ also confirmed a strong spin-lattice interaction in the system. Optical second harmonic generation (SHG) reported[10] in NiO just below $T_N$ is traced to the spin-ordering in the AFM domains and the concurrent/allied polarization in the substantial magnetic domain walls, also indicating significant ME-coupling. Weber et al.[11] established a strong magneto-elastic interaction in NiO, which governs the width and internal structure[12] of the Nèel magnetic domain walls.

$Li_{0.05}Ti_{0.02}Ni_{0.93}O$ samples were synthesized through the sol-gel method. Stoichiometric amounts of $Ni(NO_3)_2 6H_2O$, $LiNO_3$, and citric acid were mixed and dissolved into sufficient amount of distilled water, to get a clear solution. Thereafter, tetrabutyl titanate ($[CH_3(CH_2)_3O]_4Ti$) was added slowly, and the solution was heated and stirred to form the gel. Dried gel was then calcined at 800°C for 1hr. in air and finally sintered at 1100°C for 12hrs. X-ray diffraction (SXRD) patterns were collected using powder diffractometer on Indian Beamline BL-18B (Photon Factory, KEK, Japan), at X-ray wavelength of $\lambda$=1.089Å. Magnetization measurements were done on VSM-SQUID (Quantum Design) at 7Tesla fields. Dielectric measurements were collected with Alpha-A High-Performance Frequency Analyzer (NOVO-Control Technologies), using a 9T superconducting-magnet Integra System (OXFORD Instruments). Polarization loop (*P-E*) was traced at room temperature with a Precision

Premier-II ferroelectric loop tracer (Radiant Technology). Heat flow and specific heat were obtained with modulated differential scanning calorimeter MDSC-2910 (TA Instruments).

$Li_{0.05}Ti_{0.02}Ni_{0.93}O$ is synthesized in single phase, having no extra peaks to within the detection-sensitivity, as shown in fig.1 inset; confirmed by the LeBail profile peak-fitting (Full-Prof software, profile matching with constant scale factor). At room temperature, $Li_{0.05}Ti_{0.02}Ni_{0.93}O$ (LTNO) as a Li/Ti double-doped derivative has the same distorted-rhombohedral structure as the pure NiO.[7] In fig.1 (main panel) we have zoomed-in the SXRD pattern of LTNO at 250°C (fitted with cubic phase symmetry $Fm$-$3m$; lattice parameters $a = b = c = 4.1828$Å) and also at room temperature (fitted with the rhombohedral phase symmetry $R$-$3m$; lattice parameter $c = 2.9468$Å and rhombohedral angle $\alpha = 60.056°$). The (222) reflection of the high-temperature cubic phase bifurcates below $T_N = 488$K (523K for pure NiO)[13] into two reflections (202 and 006), due to the rhombohedral distortion of the lattice across the transition. $R$-$3m$ may be represented as an angular distortion of the unit cell, with a concurrent displacement of the Ni-sublattice, generating a contraction along a <111> axis of the original cubic lattice. The observed symmetry-reduction is attributed to the spin-induced internal strain, as discussed below. Rhombohedral structure in the type-II AFM state can be based on a mono-molecular unit cell, in which the angle $\alpha$ is slightly larger than 60°. In the spin-scheme used by Rooksby,[13] below $T_N$ the Ni-spins in NiO are aligned in parallel arrangement on the [111] plane and the spins on the adjacent planes are coupled antiferromagnetically with each other.[12] The same spins-scheme for the AFM ordering could be applied to the doped-NiO (LTNO), also with concomitant magneto-structural change.

A clear signature of the AFM ordering is observed in the heat-capacity (fig.2a). Heating/cooling hysteresis [($T^{**}$-$T^*$) ~7K, fig.2a inset) in the heat-flow ($W$) confirms the kinetic (first-order) character of the concurrent structural transition. Figure 2b shows $M(T)$ data of LTNO and the AFM anomaly at the applied field of 7T; nature of the transition (dip in the moment) resembles the anomaly associated with the AFM ordering in the CuO single crystals,[4] concurrent with the latter's FE-ordering. The slope- and curvature-discontinuity at $T_N$ provide ample evidence of Ni-spins' ordering. Double-doping with Li and Ti here quite suppresses the $T_N$ from 523K (pure NiO) to 488K ($Li_{0.05}Ti_{0.02}Ni_{0.93}$). Significant downshifting of $T_N$ vis-à-vis pure NiO is due both to the dilution of the magnetic-ions as well as their site-disorder. This alters the values of both $J_{NN}$ (nearest neighbor) and $J_{NNN}$ (next nearest neighbor) exchange interactions, as in the case of Mg-doped NiO.[14] The AFM ground state is also confirmed by the reversible/single-valued *M-H* curve (fig.2b inset).

Figure 3 shows the permittivity profile $\varepsilon'(T)$ of LTNO. Previously reported[5] measurement on LTNO was not recorded up to $T_N$, as the main focus was on the colossal dielectric constant (CDC) feature in the sub-$T_N$ regime. Here we present $\varepsilon^*(T)$ of LTNO up to 250°C, collected over 1Hz-5MHz. Figure 3 (left main-panel) depicts both (a) the Maxwell-Wagner (space-charge) low-frequency (~100Hz) conducting behavior (peak in $M''(\omega)$, $\varepsilon'' \sim \omega^{-1}$) due to electrical inhomogeneity and (b) the Debyean (dipolar) high-frequency (~MHz) relaxation-response of insulating background-matrix. Combined (Maxwell-Wagner + Debyean) fits (shown as solid lines) made onto the full-range spectra at 500K above $T_N$ = 488K provide hugely different characteristic timescales (~0.1s vs. ~0.1μs) for the two dielectric-mechanisms. Cole-Cole plot ($\varepsilon''$ vs. $\varepsilon'$, inset) well-illustrates these well-resolved spectral regimes, precisely switching over at 20kHz. At lower temperatures (especially at 300K), this switching-frequency reduces further. Therefore, we show in the right main-panel, permittivity $\varepsilon'(T)$ at ~MHz frequencies, as reflecting entirely *intrinsic* dipolar-characteristic. Small but clear peak-anomalies seen here apparently resemble those reported[4] in CuO for its concurrent ferroelectric (FE) transition at $T_C$ = $T_N$. The dielectric anomaly here coincides with the magneto-structural transition temperature $T_N$, implying strong coupling of the three (lattice, dipole, and spin) degrees of freedom. The anisotropic change in $\varepsilon'(T_N)$ reported[15] for MnO and MnF$_2$ (undergoing the same isostructural transitions as LTNO here) may be a possibility in the present LTNO as well.

An almost-saturating and clearly-split (hysteretic) loop observed (inset, fig.3 right-panel) indicates the presence of a finite room-temperature polarization in the AFM phase. Shape of the *P-E* loop obtained quite resembles that for the doped-BiFeO$_3$ room-temperature multiferroic,[16] latter having leaky ferroelectricity. Maximum polarization here is found to be 8000μC/m$^2$; the value almost equal to or more than those reported in the doped/pure BiFeO$_3$,[16] and in Sr$_3$Co$_2$Fe$_{24}$O$_{41}$.[17] Origin of the finite polarization in the present compound is probably rooted in the spiral pattern of the spins in the AFM domain walls. According to Mostovoy, the cycloidal-spiral spin structure in the Néel domain walls can produce an electric polarization, describable by the inverse Dzyaloshinsky-Moriya (D-M) spin-orbit interaction.[18] Due to the magneto-strictive effect, uncommonly wide (150 nm) Néel walls materialize in the pure NiO,[11] upon crystal-twinning below $T_N$. We contend that the same character of the Néel walls is carried over to our scarcely-doped $Li_{0.05}Ti_{0.02}Ni_{0.93}O$, and is responsible for its rather large electric polarization as observed.

We have directly investigated the room temperature magneto-electricity in LTNO by its permittivity measurements under magnetic fields up to 9T. Figure 4(a) left panel shows the dielectric constant vs. frequency at three different field-values over 100kHz to 4MHz (intrinsic-response) range. High-frequency permittivity $\varepsilon'(H)$ shown in the upper inset of figure 4a (left panel) features distinct 'linear' regimes, at low and high-fields. The normalized low-field linearity-prefactor $(d\ln\varepsilon'/dH)_{H\to 0}$ over the high spectral-range (shown in the lower insert) features a positive frequency-coefficient (which is but favorable), with large 20dB dynamic range. At the same key frequencies, figure 4(b) right panel shows magneto-capacitance MC($H$), evaluated as

$$\text{MC} = [\varepsilon'(H)/\varepsilon'(0) - 1] \times 100$$

The parallel ($\propto H$, log-log plot) low-field behavior of MC($H$) against frequency sustained over two decades of values signifies a robust spectral integrity of the linear magneto-electric effect. Consistent with the two asymptotic linear-behaviors evident in $\varepsilon'(H, 4MHz)$ (left panel, upper insert, as also obtained at all the relevant frequencies), MC($H$) too marks distinct low/high-field regimes. The *positive* MC obtained in LTNO may be explained by the domain-aligning effect of the field (similar[12] to that in pure NiO), which tends to abet the cycloidal rotation of the spins in the domain walls. The static polarization ($P$) pinned to the Nèel walls[11] and proportional to the spiral spin-rotation[18] ($|S_i \times S_{i+1}|$) thus reduces under the applied field, thereby enhancing the polarizability and increasing the (ac) permittivity.

In conclusion, we provide clear evidence of above room temperature magneto-electricity in $Li_{0.05}Ti_{0.02}Ni_{0.93}O$. Magnetization and high-frequency dielectric results confirm simultaneous bulk AFM ordering and an unconventional electrical organization at above-ambient-temperatures. Electrical measurements provide a rather large (8000μC/m$^2$) polarization. We establish a spectrally-robust linear magneto-electric effect, which, with positive frequency-coefficient shoots up by orders of magnitude. Polarization-character here is different from that of a long-range FE state; it is pinned to the Nèel domain-walls, and is realized by an inverse D-M interaction of the cycloidal-spiral spins within. Nèel-walls harbouring the 'polar-domains' suggest the coinage of a magnetically-induced *amorphous ferroelectric* term for LTNO as a 'peculiar room temperature multiferroic', which is easily synthesized in the single-phase and is functionally-suitable.

**Acknowledgements**

Sanjay Singh is thanked for providing XRD data, acknowledging the Department of Science and Technology (via project No. 2013-IB-013), Govt. of India. S.M. Gupta (RRCAT) is thankfully acknowledged for the *P-E* measurments.

**Figure Captions**

**Fig.1** Synchrotron XRD pattern of LTNO at room-temperature and at 250°C over zoomed-in $2\theta$ range. Inset: room-temperature full-scan SXRD data of LTNO, fitted with LeBail program.

**Fig.2** (a) Specific heat $C_p$ with peak-anomaly at the magneto-structural transition temperature and (inset) warming and cooling heat flows with 7K hysteresis. (b) Magnetic moment of LTNO at 7T in warming and cooling cycles and (inset) *unsplit* $M$ vs. $H$ upto 7T field at 350K signify a robust AFM ground state.

**Fig.3** (a) High-temperature broadband (1Hz-10MHz) spectra of permittivity ($\varepsilon^*$, left $y$-axis), modulus ($M''$, right $y$-axis), and $\varepsilon''$ vs. $\varepsilon'$ Cole-Cole plot (inset) depict the space-charge/conductive and dipoar/relaxational responses at low- and high-frequencies respectively. Solid lines-- combined Maxwell-Wagner plus Debyean fits. (b) High-frequency (~MHz) intrinsic dielectric constant, showing clear anomalies at $T_N = 214$°C, and an associated room-temperature *P-E* loop (inset) obtained at 50Hz.

**Fig.4** (a) Room-temperature intrinsic-permittivity spectra under 0, 1, and 9T fields. Upper inset depicts $\varepsilon'_{RT}(H)$/4MHz featuring two linear regimes; a sub-Tesla $d\varepsilon'/dH \sim 120\text{T}^{-1}$ and a high-field $d\varepsilon'/dH \sim 5/\text{T}^{-1}$. Lower inset: log-log plot of normalized linearity-slope $(d\ln\varepsilon'/dH)_{H\to 0}$ *increases* (+20dB change) from 0.1 MHz to 4 MHz. (b) Parallel field-scans ($\propto H_{low}$) at high-frequencies, of room-temperature magneto-capacitance signify robust spectral integrity of its linear $H$-dependence, despite its huge 20dB dynamic range over 100kHz-4MHz bandwidth.

**Figure 1**

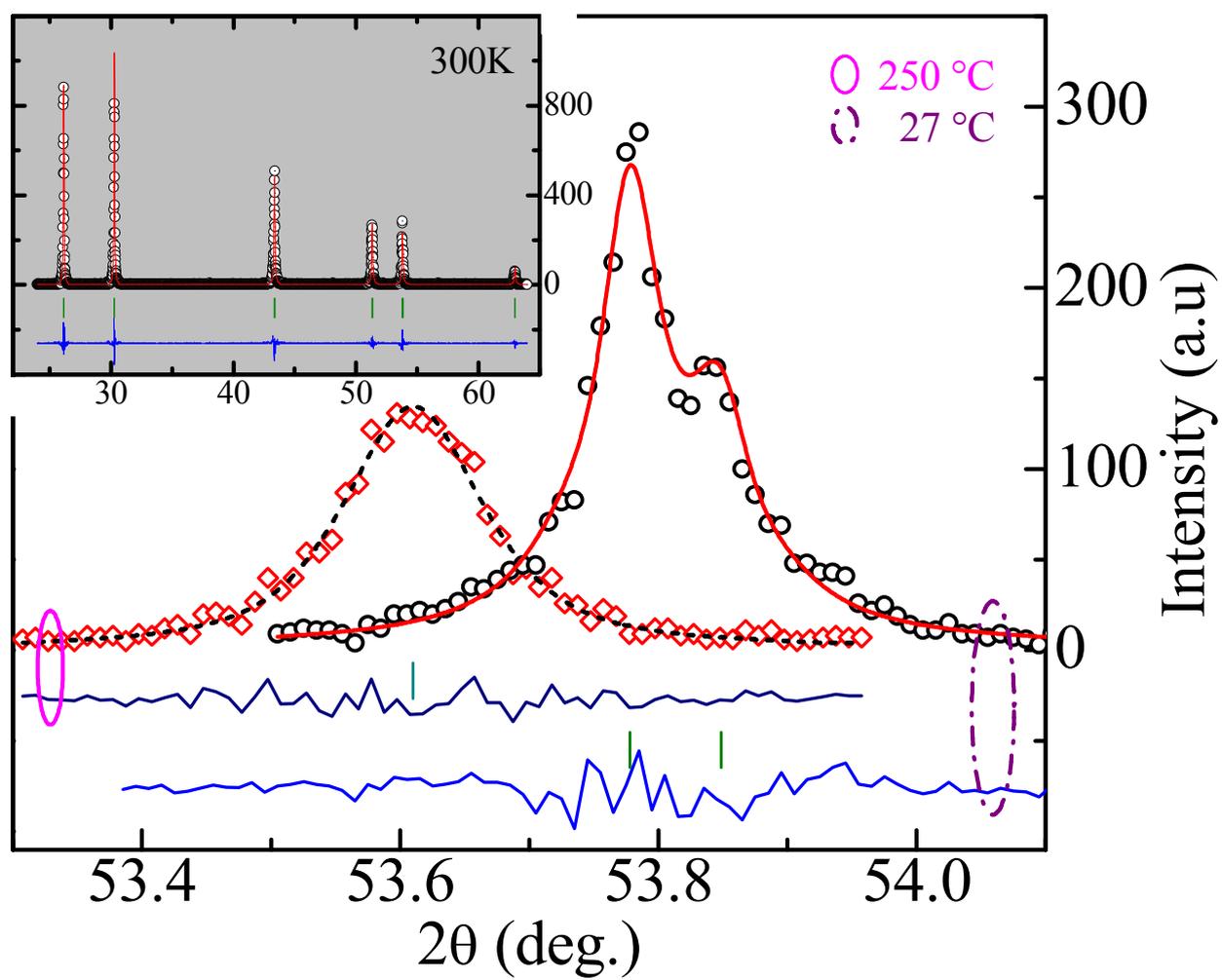

Figure 2

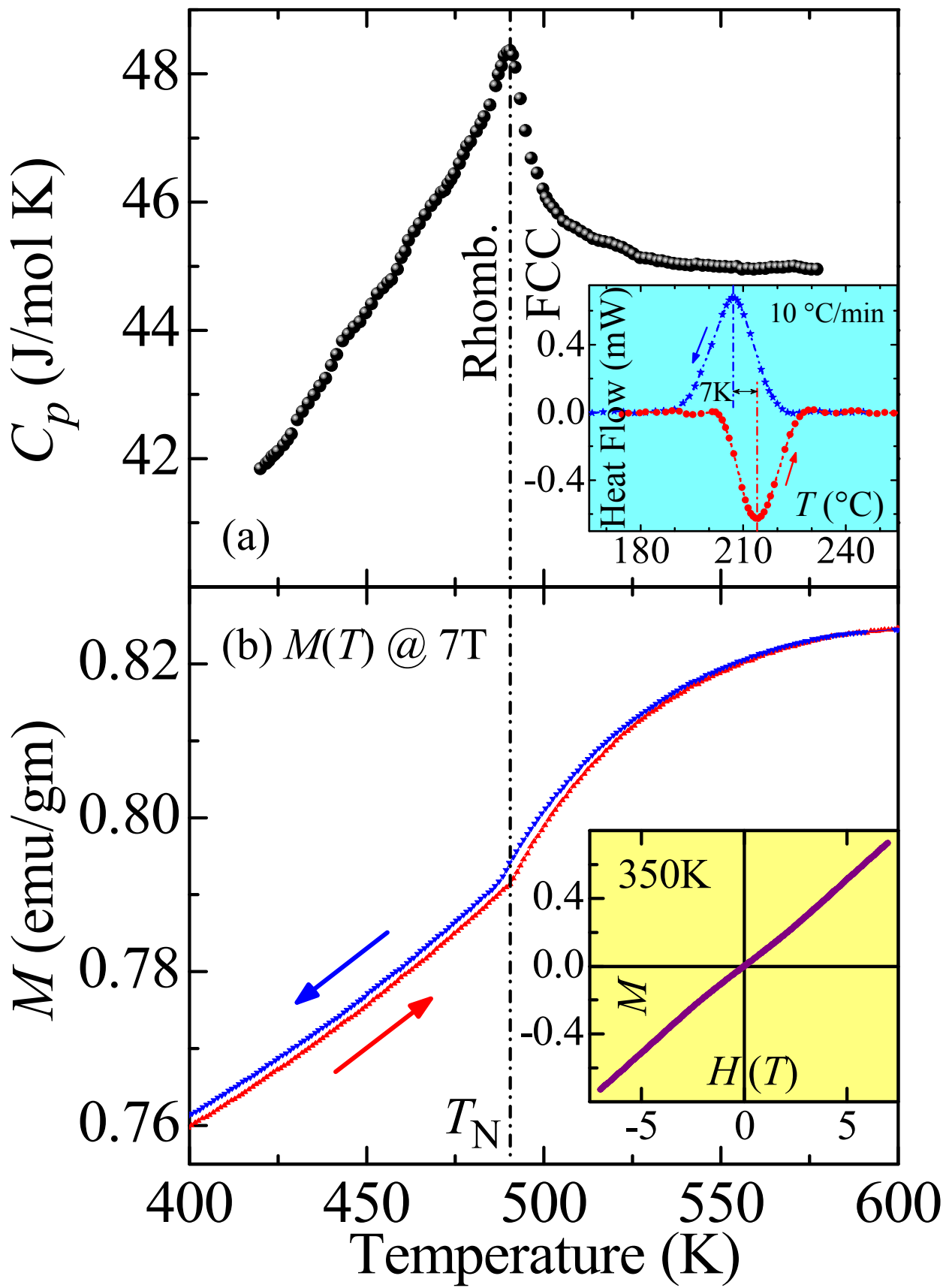

**Figure 3**

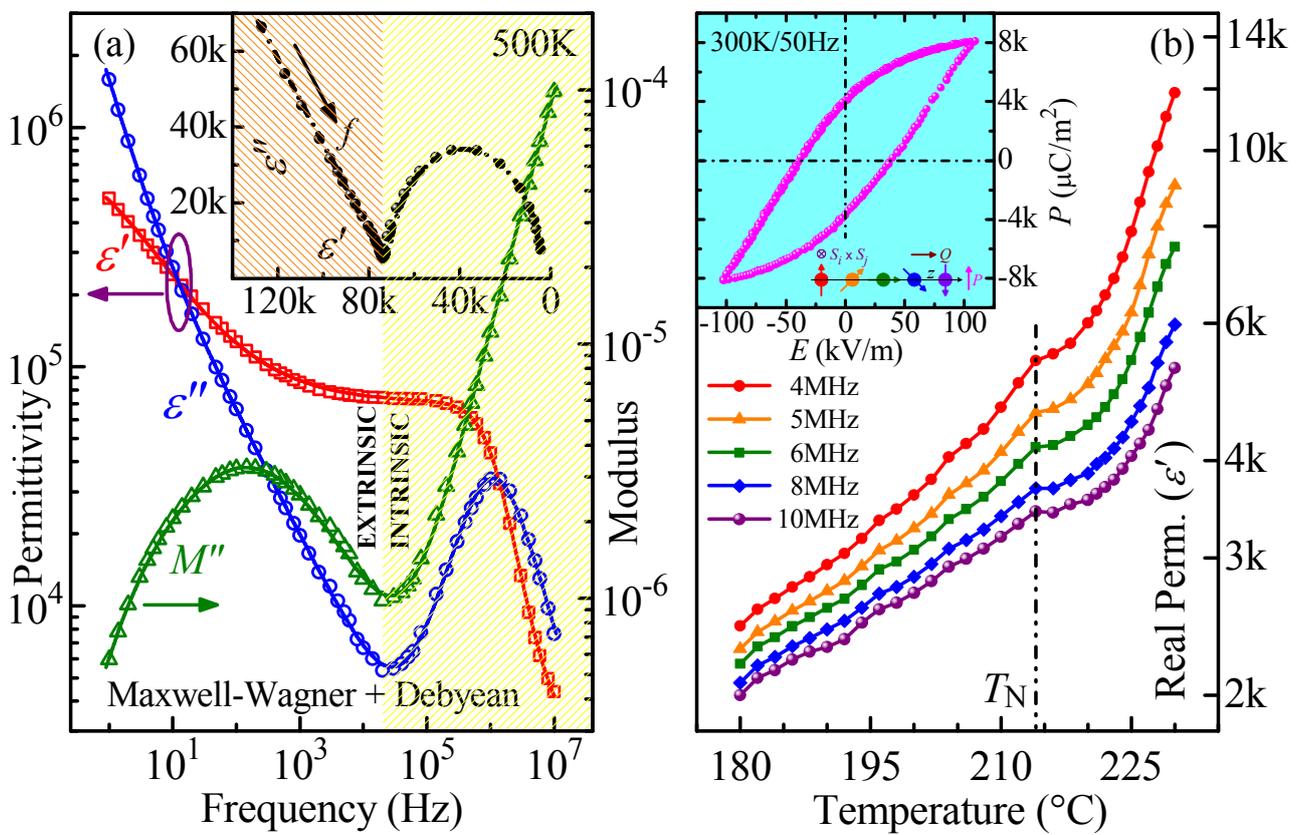

**Figure 4**

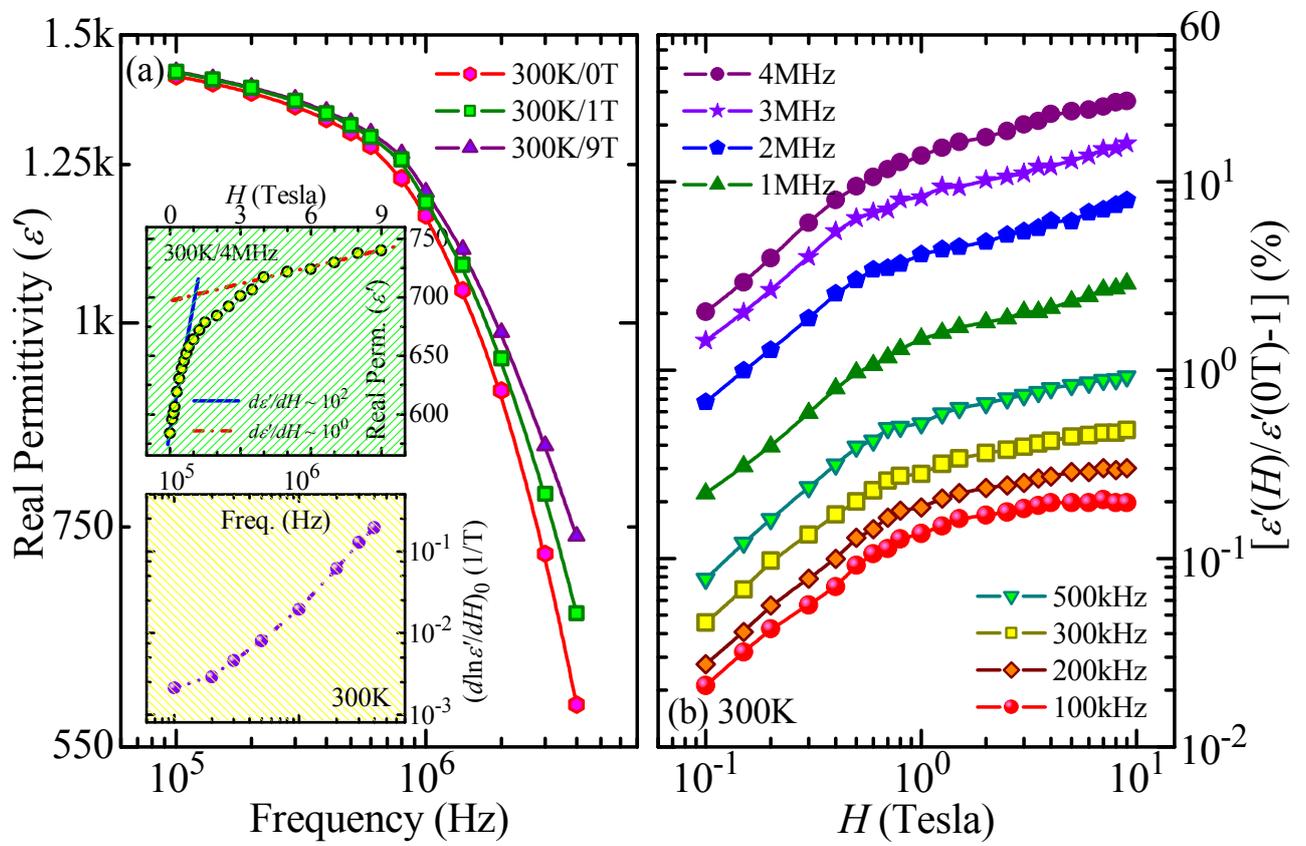